\documentclass{eas}
\usepackage{graphicx}
\pdfoutput=1
\begin{document}

\title{Asteroseismology of eclipsing binary stars using \textit{Kepler} and the HERMES spectrograph} 
\author{Valentina S. Schmid}\address{Institute of Astronomy, KU Leuven, Celestijnenlaan 200D, B -- 3001 Leuven, Belgium}
\author{J. Debosscher}\sameaddress{1}
\author{P. Degroote}\sameaddress{1}
\author{C. Aerts}\sameaddress{1}\secondaddress{Department of Astrophysics/IMAPP, Radboud University Nijmegen, P.O. Box 9010, NL -- 6500 GL Nijmegen, The Netherlands}
\begin{abstract}
We introduce our PhD project in which we focus on pulsating stars in eclipsing binaries. The combination of high-precision \textit{Kepler} photometry with high-resolution HERMES spectroscopy allows for detailed descriptions of our sample of target stars. 

We report here the detection of three false positives by radial velocity measurements.
\end{abstract}
\maketitle
\section{Introduction}
Asteroseismology provides an important tool to test and improve stellar models. It allows to deduce the interior stellar structure, such as the internal rotation profile and the extent of the core due to convective overshooting. Eclipsing binaries, on the other hand, give valuable constraints on the global fundamental parameters, like mass and radius. Thus, they provide crucial input for the seismic modeling of pulsating stars in binary systems.

We selected a sample of eight targets out of the \textit{Kepler} Eclipsing Binary catalogue (Borucki \etal, \cite{Bor10}; Slawson \etal, \cite{Slaw11}) based on the shape of the eclipses and clear pulsation-like variations in the light curves.
In 2013 we gathered high-resolution spectra with the HERMES spectrograph (Raskin \etal, \cite{Ras11}) at Mercator telescope (La Palma, Spain) to confirm the binarity.


\section{False positives}
\begin{figure} 
\includegraphics[width=0.5\textwidth]{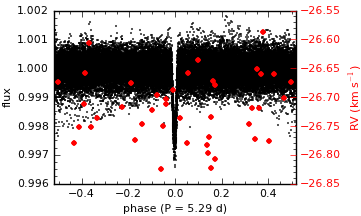} 
\includegraphics[width=0.5\textwidth]{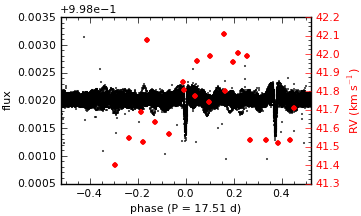} 
\caption{The \textit{Kepler} light curve (black dots) and radial velocities (red circles) of KIC3532985 (left) and KIC4565985 (right), phase folded with the periods of the eclipse signal, 5.29\,d and 17.51\,d (Slawson \etal, \cite{Slaw11}), respectively.}
\label{F_353_456}
\end{figure}

\begin{figure} 
\centering
\includegraphics[width=0.5\textwidth]{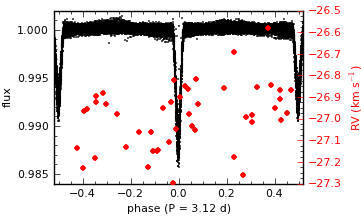} 
\caption{Same as Fig.\,\ref{F_353_456} for KIC10960993 with a period of 3.12\,d (Slawson \etal, \cite{Slaw11}).}
\label{F_109}
\end{figure}

We calculated the radial velocities by cross correlation. For three targets we did not detect any periodic variations in the radial velocity measurements (see Fig.\,\ref{F_353_456} and Fig.\,\ref{F_109}). Therefore, we conclude that these stars are not binaries. Single stars can exhibit eclipse like signals in \textit{Kepler} observations due to, e.g., background contamination or electronic crosstalk (Bryson \etal, \cite{Bry13}).

\section{Future Prospects}
As a next step we will conduct detailed studies of the pulsations and binarity of the remaining targets of our sample and aim to perform asteroseismology.

\begin{acknowledgements}
Based on observations made with the Mercator Telescope, operated on the island of La Palma by the Flemish Community, at the Spanish Observatorio del Roque de los Muchachos of the Instituto de Astrof\'isica de Canarias and obtained with the HERMES spectrograph, which is supported by the Fund for Scientific Research of Flanders (FWO), Belgium, the Research Council of KU Leuven, Belgium, the Fonds National Recherches Scientific (FNRS), Belgium, the Royal Observatory of Belgium, the Observatoire de Gen\`eve, Switzerland and the Th\"uringer Landessternwarte Tautenburg, Germany.
VSS is an Aspirant PhD fellow of the Flemish Fund for Scientific Research (FWO).
PD is a postdoctoral fellow of the Flemish Fund for Scientific Research (FWO).	
\end{acknowledgements}


\end{document}